\documentclass[useAMS,usenatbib,letterpaper]{mn2e}

\usepackage{url}
\usepackage{lscape}
\usepackage{graphicx}
\usepackage{enumerate}
\usepackage{float}
\usepackage{floatflt}
\usepackage{fancyvrb}
\usepackage{epstopdf}

\usepackage{pdfpages} 
\title{\textit{pt5m} - a 0.5m robotic telescope on La Palma}
\author{L. K. Hardy,$^{1}$ T. Butterley,$^{2}$ V. S. Dhillon,$^{1,3}$ S. P. Littlefair,$^{1}$ R. W. Wilson,$^{2}$
\\
$^{1}$Department of Physics and Astronomy, University of Sheffield, Sheffield, S3 7RH, UK\\
$^{2}$Centre for Advanced Instrumentation, Department of Physics, University of Durham, South Road, Durham, DH1 3LE, UK\\
$^{3}$Instituto de Astrof\`{i}sica de Canarias, E-38205 La Laguna, Tenerife, Spain}

\begin{document}

\date{September 2015}

\pagerange{\pageref{firstpage}--\pageref{lastpage}} \pubyear{2015}

\maketitle

\label{firstpage}

\begin{abstract}
\textit{pt5m} is a 0.5m robotic telescope located on the roof of the 4.2m William Herschel Telescope (WHT) building, at the Roque de los Muchachos Observatory, La Palma. Using a 5-position filter wheel and CCD detector, and bespoke control software, \textit{pt5m} provides a high quality robotic observing facility. The telescope first began robotic observing in 2012, and is now contributing to transient follow-up and time-resolved astronomical studies. In this paper we present the scientific motivation behind \textit{pt5m}, as well as the specifications and unique features of the facility. We also present an example of the science we have performed with \textit{pt5m}, where we measure the radius of the transiting exoplanet WASP-33b. We find a planetary radius of $1.603\pm 0.014 R_{J}$.
\end{abstract}

\begin{keywords}
instrumentation: miscellaneous, photometers -- telescopes -- transients -- planetary systems -- stars: individual: WASP-33.
\end{keywords}


\section{Introduction}\label{sec:intro}
Robotic observing facilities requiring no human input during night time are not new or uncommon\footnote{For a global list of robotic telescopes see \url{http://www.astro.physik.uni-goettingen.de/~hessman/MONET/links.html}}, and are becoming ever more popular (e.g. \citealt{zerbi01,french04,tsapras09,gillon11,brown13,gorbovskoy13,andersen14,schmitt14,strogler14}). The largest example is the 2m Liverpool Telescope \citep{steele04}, providing a fully autonomous, large aperture telescope at a high quality site, with a suite of instrumentation available. Such facilities can sharply reduce operation costs as they require much less staff interaction. They can also be programmed to make complex scheduling decisions in order to observe the most appropriate targets, given the observing conditions. In addition, using automated transient alert-listening software, they can respond to unexpected events much faster than human-controlled telescopes.

\begin{figure}
\centering
 \includegraphics[width=0.48\textwidth]{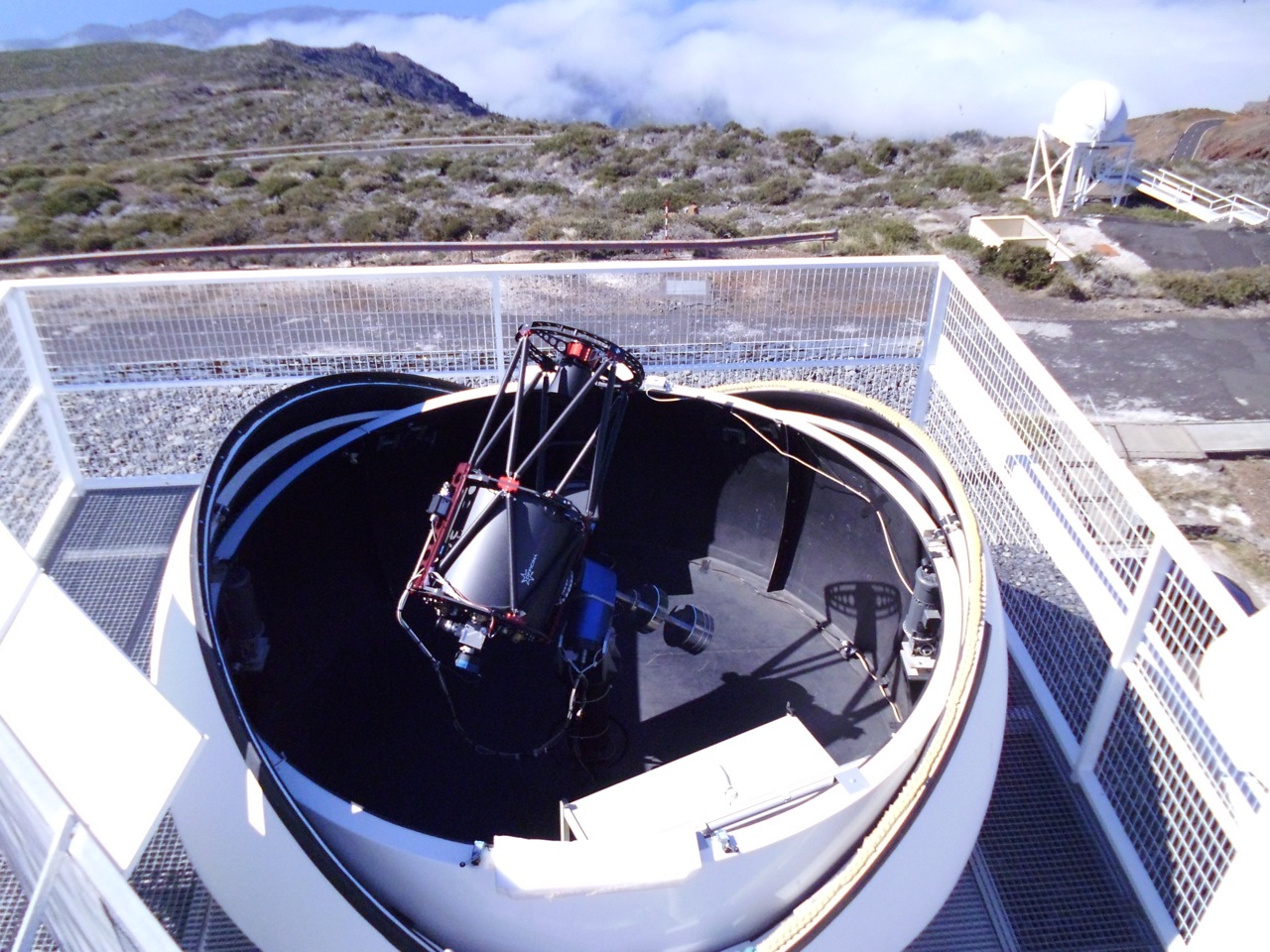}
 \caption{Photograph of \textit{pt5m} on the roof of the WHT taken in April 2015. The WHT dome is out of view to the left.}
\label{fig:photo}
\end{figure}

\textit{pt5m}\footnote{The name is an abbreviated form of `point-five-metre'} is a 0.5m robotic telescope located on the roof of the WHT on La Palma (Figure \ref{fig:photo}) and is hosted by the Isaac Newton Group of Telescopes (ING). It functions as an astrophysics  research facility, an atmospheric turbulence profiler, and a teaching resource. \textit{pt5m} has already been used to study pulsating stars, transiting exoplanets, solar system objects, interacting binaries \citep{littlefair13,kupfer14,campbell15b}, and transients \citep{gandhi14,hardy15atel}, but no detailed description of \textit{pt5m} itself has appeared in the refereed astronomical literature to date. In this paper we therefore detail the hardware and software architectures (Sections~\ref{sec:hardware}~\&~\ref{sec:software}), the telescope performance (Section \ref{sec:comms}), and an example photometric study of the transiting exoplanet WASP-33b (Section \ref{sec:results}).

\begin{figure}
\centering
 \includegraphics[width=0.48\textwidth]{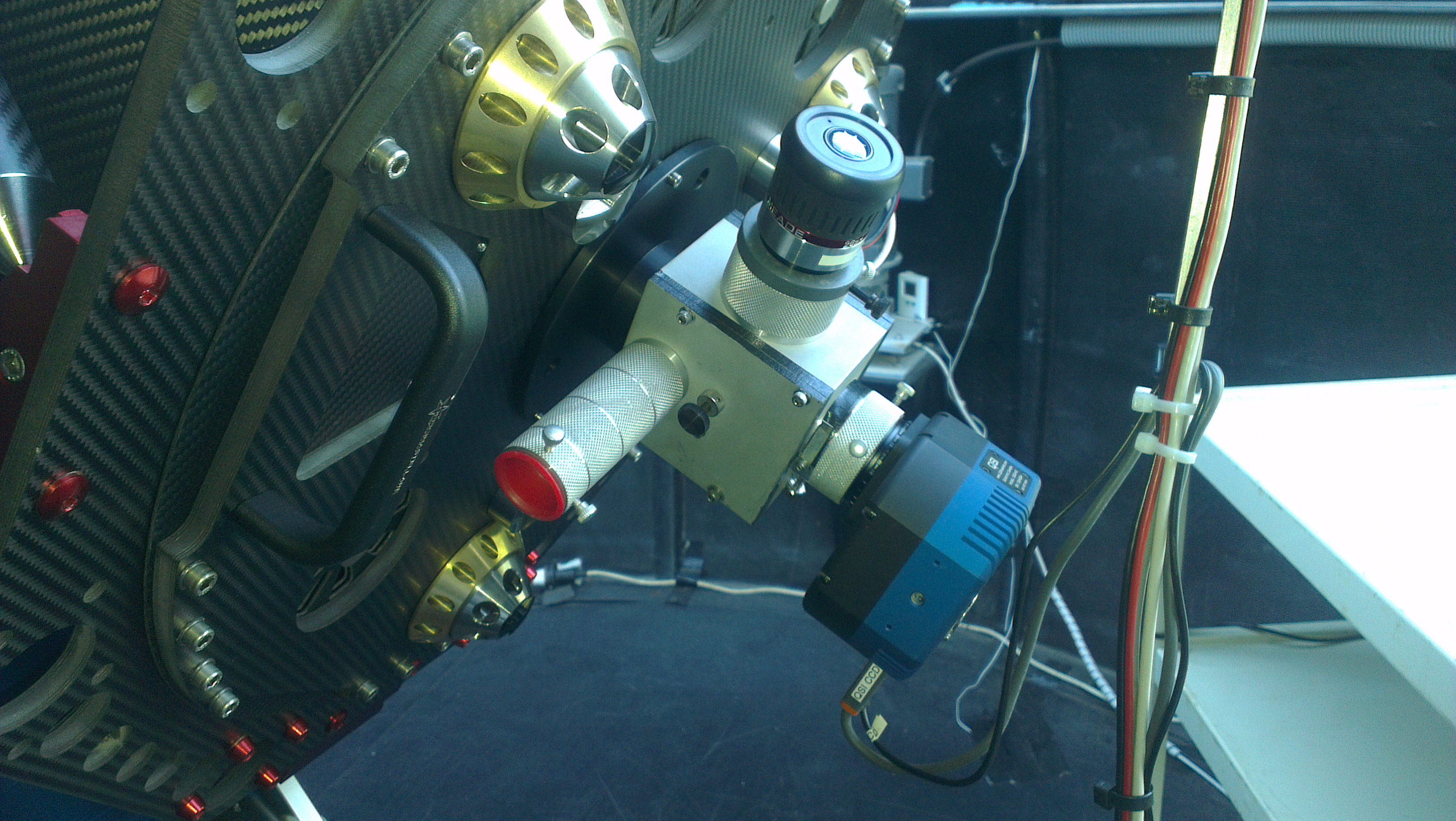}
 \caption{Photograph of the Cassegrain focus of \textit{pt5m}, showing the flip mirror mechanism (grey), and the CCD camera (blue) with integral filter wheel (black). In this photo the SLODAR port (pointing upwards) temporarily holds an eyepiece.}
 \label{fig:backend}
\end{figure}


\section{Brief History of \textit{\MakeLowercase{pt5m}}}\label{sec:history}
\textit{pt5m} was first developed as an atmospheric turbulence profiling (SLODAR) facility \citep{wilson02} and was originally installed at the South African Astronomical Observatory \citep{catala13}. The telescope was then moved to La Palma in 2010 to support the CANARY adaptive optics demonstrator \citep{morris09}. For this reason it was sited as close as possible to the WHT, to sample the turbulence in approximately the same beam as the WHT. 

In order to fully utilise the telescope when not being used for SLODAR observations, we decided to add a science imager in 2011. We recognised that even a small telescope located at such an excellent site could provide a powerful tool for time-domain astronomy. The science CCD and integrated filter wheel (see Section \ref{sec:qsi}) were originally mounted on a linear slide to move the science camera in and out of the light path of the SLODAR optics. By 2012 we had implemented the robotic mode (Section \ref{sec:software}) and started using \textit{pt5m} on a routine basis for time-resolved and transient astronomy. In 2014 the original 0.5m Orion Optics Modified Dall-Kirkham telescope was replaced with a new Ritchey-Chretien telescope (Section \ref{sec:tel}). This refurbishment was made to improve throughput and image quality, which were not a high priority for the original SLODAR studies. At the same time the slide mechanism was removed and replaced with a flip mirror that allows light to be diverted to either one of two optical ports. One of these ports hosts the science camera and the other hosts the atmospheric turbulence profiler (see Figure \ref{fig:backend} and Section \ref{sec:tel}).


\section{Hardware}\label{sec:hardware}
\begin{figure}
\centering
 \includegraphics[width=0.48\textwidth]{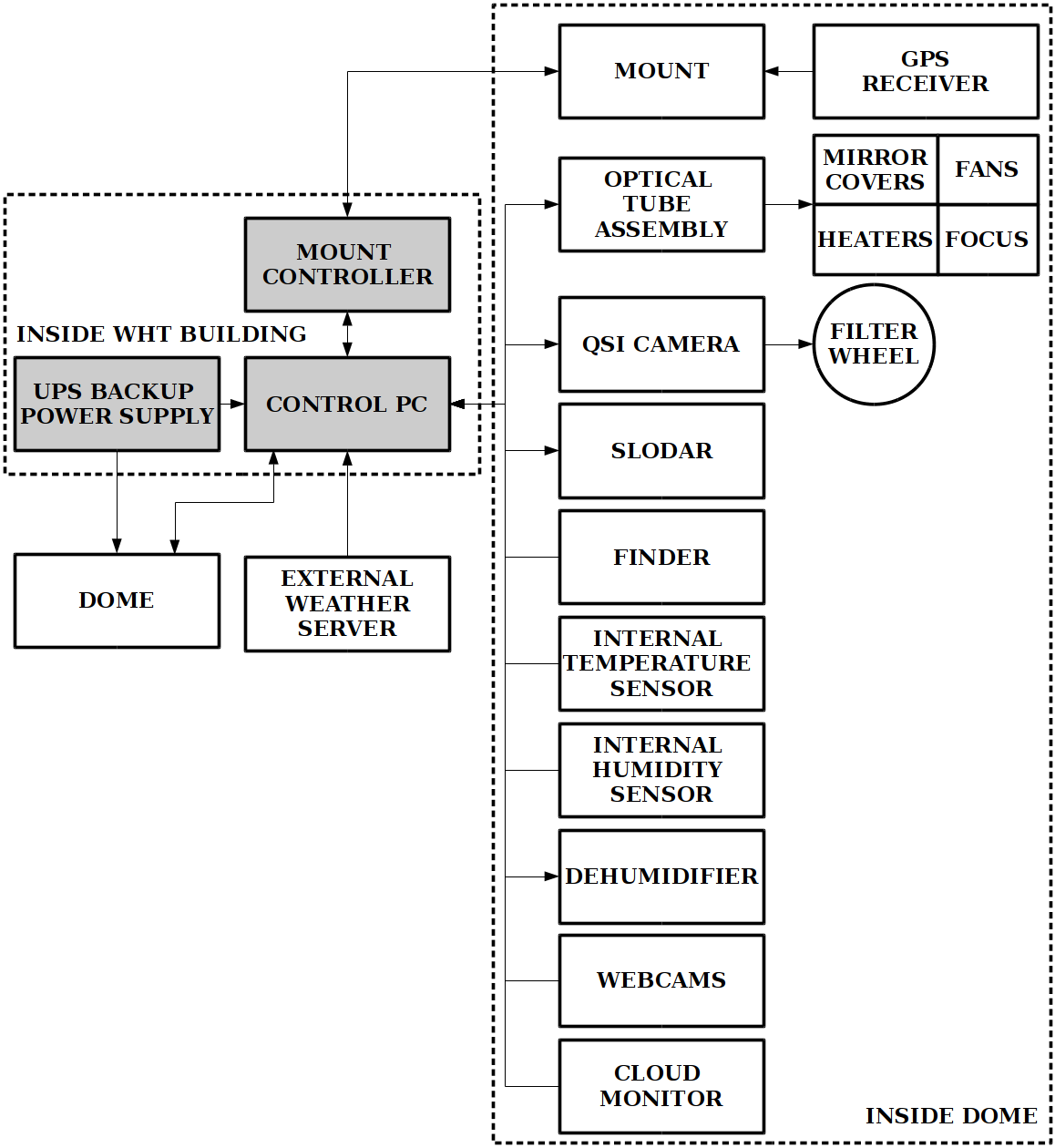}
 \caption{Schematic showing the \textit{pt5m} hardware components and their connections.}
 \label{fig:hardware}
\end{figure}

The following subsections describe the hardware components of \textit{pt5m}. Figure \ref{fig:hardware} shows the components schematically, whilst Table \ref{tab:parameters} summarises the facility specifications.

\begin{table}
\begin{center}
\caption{Specifications of \textit{pt5m}. Location information is as measured by GPS (WGS84).}
\label{tab:parameters}
\begin{tabular}{p{4.8cm} p{2.9cm}}
  \hline
  \textbf{Parameter} & \textbf{Value}  \\ \hline
  \textbf{Telescope} \\ 
  Longitude & 28$^{\circ}$ 45' 38.7" N \\ 
  Latitude & 17$^{\circ}$ 52' 53.2" W \\ 
  Altitude & $2383\pm3$m \\ 
  Primary Diameter & 0.5m \\ 
  Primary Focal Length & 1500mm \\
  Primary Conic Constant & -1.071 \\
  Secondary Diameter & 0.148m \\ 
  Secondary Focal Length & -609.89mm \\
  Secondary Conic Constant & -4.182 \\
  Dome Diameter & 3.66m \\ 
  Focal Length & 5000mm \\ 
  Platescale & 41.25"/mm \\
  Telescope un-vignetted FoV & 9' \\ 
  Slew Speed & 20$^{\circ}$/second \\ 
  Pointing Accuracy (RMS) & 0.4' \\ 
  Tracking Accuracy & 12"/hour \\ 
  Finder Field of View & $7.2^{\circ}\times 5.4^{\circ}$  \\ \hline
  \textbf{Instrument} \\ 
  Detector Format & 2184 $\times$ 1472 pixels \\ 
  Pixel Size & 6.8 $\mu$m \\ 
  Pixel Scale & 0.28"/pixel \\ 
  CCD Field of View & 10.2' $\times$ 6.9' \\ 
  Full Well Depth & 83000 e$^{-}$ \\ 
  Readout Noise (binned 1x1) & 10.4 e$^{-}$ \\ 
  Readout Noise (binned 2x2) & 14.4 e$^{-}$ \\ 
  Readout Noise (binned 3x3) & 18.5 e$^{-}$ \\ 
  Gain & 1.3 e$^{-}$/ADU \\ 
  Flat field noise (grain) & $0.15\pm0.05$\% \\
  Dark Current & $8\pm4$ e$^{-}$/pixel/hour \\ 
  Dead Time (binned 1x1) & 8 seconds \\ 
  Dead Time (binned 2x2) & 5 seconds \\ 
  B-band Zeropoint & 21.91 mag \\ 
  V-band Zeropoint & 22.06 mag \\ 
  R-band Zeropoint & 22.16 mag \\ 
  I-band Zeropoint & 21.05 mag \\ 
 \hline  
\end{tabular}
\end{center}
\end{table}

\subsection{Dome and Mount}
\textit{pt5m} is enclosed in a 12-foot Astrohaven clamshell dome. The use of a clamshell design is vital for rapid follow-up work, as there is no delay whilst the dome rotates to catch up with the telescope.

The telescope mount, a direct-drive Astelco Systems NTM-500, is of German-Equatorial design. This mount design requires a `pier-flip' when tracking a target across the observer's meridian from East to West. The pier flip introduces a discontinuity in the $x$-axis of time-series observations of up to several minutes, as the mount must wait for the target to pass through the observer's meridian before reacquiring. The pier flip also rotates the field of view through $180^{\circ}$, placing the target and comparison stars on different parts of the chip, with different levels of vignetting. This results in a small step in the differential photometry which, in principle, can be corrected for with flat fields. The only major negative effect of the pier flip is when it occurs in the middle of an interesting event, e.g. during an eclipse.

In its implementation in \textit{pt5m}, the mount has an RMS pointing accuracy of $\sigma=0.4$' around the mean pointing position, and an average tracking drift of 12"/hour. These values are larger than expected from the manufacturer's specification, most probably due to flexure in the telescope and roof-top foundations, and small errors in the polar alignment.

Although it is stable during normal use, the telescope and mount both suffer from vibrations in high winds (8-10 m/s), as the clamshell dome leaves them exposed. They also vibrate when the main WHT dome motors are engaged, as the facility's roof-top foundations are not fully isolated from the building. We are in the process of erecting substantial wind baffling around the dome, particularly on the South-West side where the wind is forced between the WHT dome and the rest of the building.

An advantage of this particular mount is its fast slewing speed of 20$^{\circ}$/second\footnote{In fact, with a smaller payload, the mount can slew at up to 100$^{\circ}$/second.}. This allows new targets to be acquired very quickly, which is essential for rapid follow-up work, as well as reducing overheads between observations. 

\subsection{Science Camera}\label{sec:qsi}
The `Quantum Science Imaging' QSI-532ws camera uses a Kodak KAF-3200ME CCD detector. The science camera is equipped with an integral 5-position filter wheel, which at present holds the following Astrodon filters: Johnson $B$, Johnson $V$, Cousins $R$, Cousins $I$, and H$\alpha$ 5nm. The camera employs a thermoelectric cooling system, and is maintained at $-20^{\circ}$C when in use. The CCD specifications are listed in Table \ref{tab:parameters}. Of particular note is the low dead-time of the CCD (5 seconds in 2x2 binned mode), which reduces overheads and improves the cadence of time-series observations. The CCD camera uses a shutter, and takes timing information directly from the control PC (see Section \ref{sec:electronics}). The timestamps are moderated by the Network Timing Protocol (NTP) which has a typical accuracy of $<100$ms. As well as a science camera, \textit{pt5m} also has an ASTROVID StellaCam finder camera mounted on the telescope.

\subsection{Telescope}\label{sec:tel}
The 0.5m optical tube assembly (OTA) is a custom built $f$/10 Officina Stellaire Pro RC 500, made almost entirely of carbon fibre, resulting in a light-weight, rigid and low thermal-expansion structure.
The OTA includes primary mirror covers, primary mirror heaters to prevent the formation of dew (these are not used), and fans for cooling the primary mirror. The telescope focus is adjusted by moving the motorised secondary mirror. 

A tertiary mirror is housed in a \textit{Van Slyke Instruments Flipper} (Figure \ref{fig:backend}). It uses a mirror on a hinged support that flips up and down. This allows the light beam to either pass straight through to the filter wheel and CCD or to be folded by 90$^{\circ}$ into the port dedicated to atmospheric turbulence profiling.

\subsection{Weather System}
In order to keep \textit{pt5m} safe from weather damage, a reliable weather station, with backup systems available in case of failure, is essential. Since \textit{pt5m} is hosted by the ING, we are able to benefit from access to their weather stations\footnote{\url{http://catserver.ing.iac.es/weather/index.php}}. These include the main weather system for the whole observatory located on a mast close to the Jacobus Kapteyn Telescope (JKT), as well as a local mast located on the WHT. In addition, \textit{pt5m} employs its own internal sensors, including a humidity and temperature sensor located in the dome. The Conditions Monitor (see Section \ref{sec:conditionsmonitor}) monitors the weather information and decides when the dome should be open or closed. The dome is forcibly closed if this weather data becomes unavailable, or returns exactly the same values for a period of more than 20 minutes (we assume this only happens if the ING weather server has crashed and is not supplying reliable data). 

A dehumidifier has been installed in the dome, which turns on automatically when the humidity in the dome is above the safe observing limit. This ensures the unit is only engaged when the dome is closed, and maintains the internal humidity at a safe level when the external humidity is high.

The dome is also equipped with an AAG CloudWatcher cloud detector. This is mounted on the internal rim of the dome and sees the entire sky when the dome is open. The unit uses an infrared sensor to measure the temperature of the sky. We have calibrated this sensor against images taken by the Gran Telescopio Canarias all-sky camera over the course of a year. We found a linear relationship between the infrared sky temperature and cloud coverage in oktas, where 0 represents a completely clear sky and 8 is completely overcast. We found no evidence for seasonal variations. We use this cloud coverage data as a guide to determine whether or not a night is photometric, although we do not currently use this information when scheduling observations robotically. Since the cloud monitor is mounted on the inside rim of the dome, it cannot be used to make a decision on whether or not to open or close the dome.

\subsection{Electronics}\label{sec:electronics}
All of the \textit{pt5m} hardware is controlled by a dedicated Linux PC on site. The machine serves as an interface between the different information inputs (e.g. weather data) and outputs (e.g. dome control). Both the control PC and the dome motors are connected to an uninterruptible power supply (UPS), so that the dome can always close even in the event of a power failure. All of the \textit{pt5m} hardware components are connected to remote-control power switches, enabling them to be turned on and off over the network. The control PC, UPS, mount controller and remote power switches 
are all located inside a purpose-built enclosure in the WHT elevator control room, instead of inside the dome. By locating the electronics inside the WHT building, we reduce the possibility of damage by sudden adverse weather conditions.

\textit{pt5m} is also equipped with three physical emergency stop buttons. The first stops the telescope and mount, engaging the brakes immediately to stop any potential damage if the telescope is about to crash into the pier, for example. This button is mounted on the telescope mount. The second button is mounted on the interior dome wall, near to the access hatch. This button shuts off power to the dome motors to prevent accidents when climbing through the hatch. The third button is located inside the WHT elevator control room, and is used to shut down the telescope and close the dome in the event of software malfunction or severed connectivity to the external network. 


\section{Software}\label{sec:software}

All of the software used to control \textit{pt5m} is written in Python and runs on the Linux control PC (Section \ref{sec:electronics}). \textit{pt5m} can be used in either ``manual" or ``robotic" mode.
In manual mode, a human user controls the facility in real time by sending a series of instructions via a command-line terminal that has been opened on the control PC. This mode is used for testing
and teaching purposes, and can be used to control \textit{pt5m} locally or remotely over the internet. In robotic mode, \textit{pt5m} operates autonomously, opening and closing the dome, monitoring the weather, taking and processing calibration data, and selecting and observing science targets.

\begin{figure}
\centering
 \includegraphics[width=0.48\textwidth]{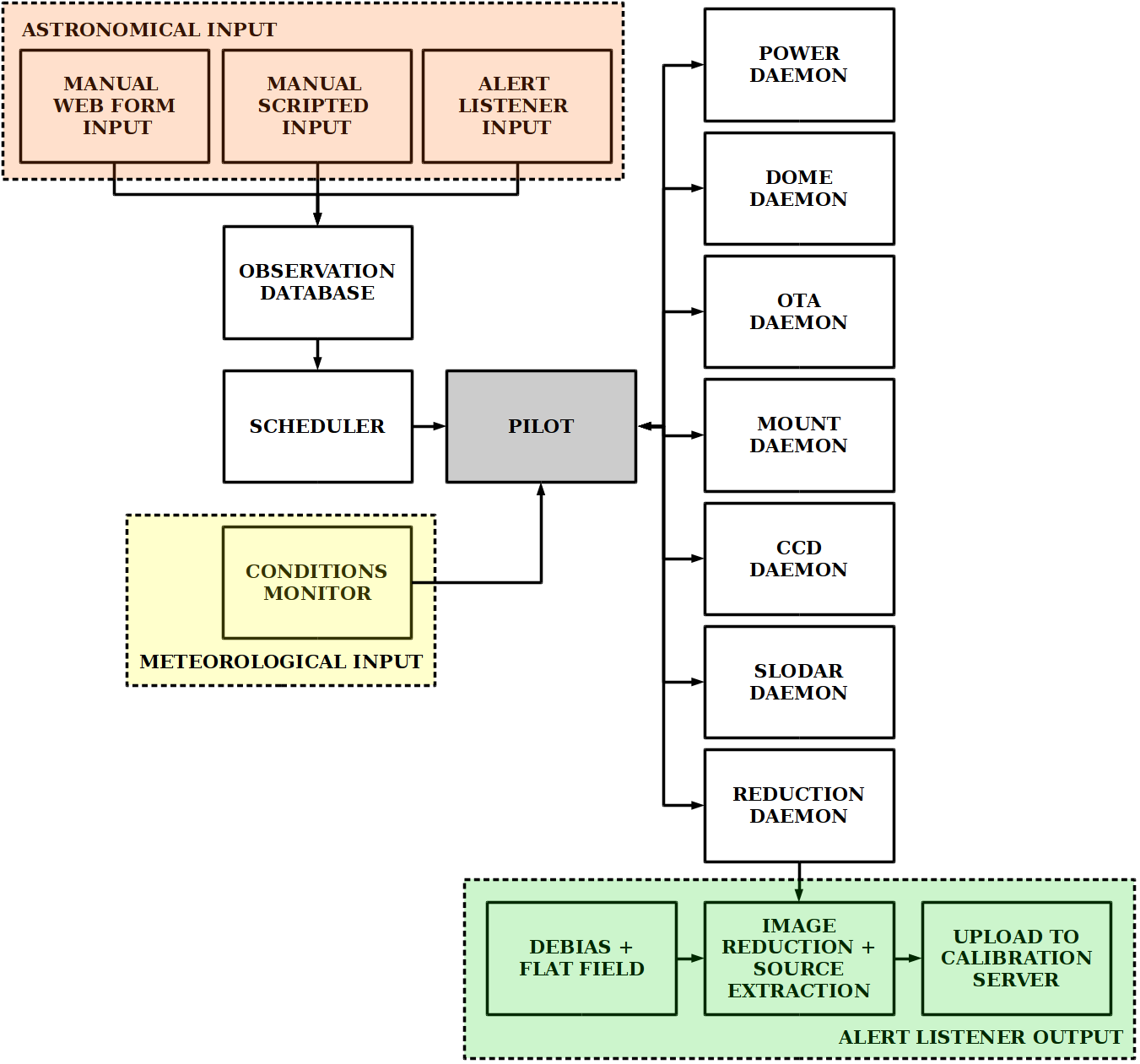}
 \caption{Schematic showing the software components of \textit{pt5m} and their interaction.}
 \label{fig:software}
\end{figure}

A schematic showing the interactions between the \textit{pt5m} software components is shown in Figure \ref{fig:software}.

\subsection{Hardware Control: Daemons}\label{sec:daemons}
The main hardware components of the system each have a dedicated software process or ``daemon" that runs in the background. These daemons allow the hardware components to be initialised independently from one another. The ``Python Remote Objects" (Pyro) library is used for communication between the daemons and other software processes -- each daemon is a Pyro server. This server/client model also allows multiple software components to query and send commands to a piece of hardware without clashes and without having to wait for a response.

The \textit{pt5m} hardware daemons are as follows:

\begin{itemize}
\item Power: Communicates with the remote-controlled power switches (Section \ref{sec:electronics}), used to power the individual hardware components on and off. This daemon runs at all times.
\item Dome: Controls opening and closing of the dome via an Arduino interface. It constantly checks the current weather conditions via the Conditions Monitor (Section \ref{sec:conditionsmonitor}). It will automatically close the dome (or refuse to open it) if any of the Conditions Monitor flags are set to 1, if the external power fails or if the network connection to the outside world fails. This daemon runs at all times.
\item OTA: Controls the OTA's built-in electronics via an RS232 interface. Functionality includes opening/closing the primary mirror cover, enabling/disabling the mirror heaters, and focusing.
\item Mount: Communicates with the telescope mount controller via an internal network. This daemon controls the pointing and tracking of the telescope. It also constantly monitors a wide range of status information that is available from the mount and makes this available to other software processes.
\item CCD: Communicates with the CCD camera via a USB interface. While the daemon is running the thermoelectric cooler is enabled and the CCD is ready to take images. The daemon maintains a queue of requested exposures (each with its own filter, exposure time and binning settings) which are processed sequentially. When this daemon is shut down at the end of each night the CCD temperature rises in a controlled manner before the CCD is powered off.
\end{itemize}

\subsection{System Control: Pilot}\label{sec:pilot}
The ``pilot" program is the software process that is responsible for autonomous control of \textit{pt5m} in robotic mode. It gathers information from all of the daemons and other processes, and issues commands to them. The pilot is started automatically each day in the late afternoon. It remains dormant during daytime, and begins preparing for the night's observations during twilight.

The evening-twilight startup procedure begins when the Sun reaches 10$^{\circ}$ above the horizon, and is as follows:
\begin{enumerate}[\bf1.]
\item Sun at 10$^{\circ}$: power on CCD and mount.
\item Sun at 7$^{\circ}$: take bias frames (dome still closed).
\item Bias frames complete; open dome and then open mirror covers.
\item Sun at 5$^{\circ}$: point telescope to a blank field in the sky and take flats.
\item Focus the telescope using a bright star.
\item Process flats and biasses into master frames if enough individual frames were taken in good conditions, i.e. no cloud presence detected during flat field observations.
\item Optionally, observe flux standards and collect dark frames. 
\item Park the telescope and wait for the end of twilight. 
\end{enumerate}
Focussing is conducted automatically on a bright Gliese catalogue star. The half-flux diameter (the radius at which half of the total encircled energy is contained) is measured at the current focus position, and then measured again at a position several millimetres away from the original position of the secondary mirror. The behaviour of the half-flux diameter away from the best focus is linear with respect to the position of the secondary mirror (focus). These two measurements, along with a predetermined function for the half-flux diameter behaviour, which does not vary with time, are all that is required to calculate the best focus position. If the autofocussing program fails (because a bright star could not be found, or the half-flux diameter measurements do not behave as expected), the telescope returns to the focus used on the previous night. 

The pilot is considered to be in the night-time state while the Sun is below $-10^{\circ}$ altitude. During this time, its behaviour is governed by the ``scheduler" (see Section \ref{sec:scheduler}). The pilot periodically queries the scheduler and, if the scheduler suggests a new target should be observed, the pilot slews the telescope and adds the specified exposures to the CCD queue. Once the pilot verifies that the telescope is tracking on the target, the CCD exposures begin. The system then remains in this state, taking exposures, until one of the following occurs:
\begin{itemize}
\item all exposures in the CCD queue have been taken and the observation is complete,
\item the scheduler returns a new, higher priority target to observe,
\item conditions force the dome to close, or 
\item the end of the night is reached.
\end{itemize}
If the scheduler returns no valid targets, the telescope is moved to the park position. However, the dome remains open so that the system is ready to respond immediately when a new target is requested. 

In the event of bad weather, the dome is closed immediately by the dome daemon (see Section \ref{sec:daemons}) without the intervention of the pilot. This is an essential precaution to protect the instrumentation from wind and water damage as the weather conditions can change quickly. The pilot then responds as follows: if the telescope is currently observing a valid target it will pause the queue of CCD exposures and leave the telescope tracking. If the conditions improve after a short time then the pilot simply reopens the dome and resumes the CCD exposures. If/when the current observation becomes invalid (e.g. the target sets below the horizon), the pilot parks the telescope and clears the CCD queue.

The morning-twilight shutdown procedure is:
\begin{enumerate}[\bf1.]
\item Sun at $-9^{\circ}$: point telescope to blank field and take flats.
\item Sun at 0$^{\circ}$: park telescope, close mirror covers, close dome.
\item Power off mount and CCD.
\end{enumerate}

\subsection{The Scheduler}\label{sec:scheduler}
The scheduler is a Python script that is interrogated every 15 seconds by the pilot during the night. The script returns a decision on what should be done, either to continue observing what is currently being observed for the next 15 seconds, or to move immediately to a new target.

Each new target (also referred to as a `job' or `pointing') is described by an ASCII text file that defines the position of the target, the CCD exposures required, the priority (an integer from 0 to 5, where 0 is the highest priority), and various other constraints on the required observations, as shown in Figure \ref{fig:jobfile}. Only one target per job is allowed. Each job is given a unique ID number, which is also the name of the ASCII file. The jobs define the `queue', and all of the awaiting job files are stored in a single directory on the control PC. When a job is complete, the file is moved from the queue directory to a `completed' directory. Alternatively, if the job is aborted, deleted, expired or interrupted, it is moved to a correspondingly named directory. 

\begin{figure*}
\begin{minipage}{1.\textwidth}
{\fontsize{1.9mm}{0.75em}\selectfont
\begin{Verbatim}[frame=single]
# Pointing     Object   RA(J2000)  Dec(J2000)  Priority  ToO  Flip  Sun Alt (deg)  Min.Alt.(deg)  Min.Time(sec)  Max.Moon  User        StartUTC             StopUTC           guide
#--------------------------------------------------------------------------------------------------------------------------------------------------------------------------------------
  11490      ASASSN14gu  104.032   33.6016        4       0    1       -15.0          35             9000          B        lkh   2015-02-19T18:00:00   2015-02-20T08:00:00     1
#
#
#
#TYPE    FILTER   BINNING   EXPTIME   NUMEXP
#--------------------------------------------
SCIENCE     V        2        150       200
DARK                 1        150         5
\end{Verbatim}
}
\caption{Example ASCII text file defining a pointing. The first variable is the unique pointing ID, the priority ranges from 0 (highest priority) to 5 (lowest priority), ToO = Target of Opportunity and is a flag used for targets that need override status, Flip is a flag used to indicate whether the observations can tolerate a pier flip, Sun Alt is the maximum acceptable altitude of the Sun, Min.Alt. is the minimum acceptable altitude of the target, Min.Time is the minimum amount of time the telescope should spend on the observations in order for them to be useful, Max.Moon takes values $B$, $G$ or $D$ for bright, grey or dark time, StartUTC and StopUTC define the time period in which the observations may begin, but once initiated they can extend beyond StopUTC, and guide is a flag used to indicate whether autoguiding should be used. The pointing file also defines the exposures requested, showing the filter, binning mode, exposure time and the number of exposures. The different types of exposures available are SCIENCE, FOCUS, DARK, SKY and BIAS.}
\label{fig:jobfile}
\end{minipage}
\end{figure*}

\begin{table}
\begin{center}
\caption{Flags used by the scheduler. Each flag can only take the value 0 (legal) or 1 (illegal). The artificial horizon refers to a map of minimum observable altitude for any given azimuth, which ensures that \textit{pt5m} never tries to observe targets which are vignetted by the WHT dome or building. See Figure \ref{fig:jobfile} for an explanation of the terms StartUTC, StopUTC, Sun Alt, Min.Alt. and Min.Time.}
\label{tab:schedulerFlags}
\begin{tabular}{p{1.55cm} p{6.2cm}}
  \hline
  \textbf{Flag name} & \textbf{Flag=0 if:}  \\ \hline
  time & The time now is between StartUTC and StopUTC \\ 
  minalt & The target is above Min.Alt. now\\ 
  mintime & The target will still be above Min.Alt. after Min.Time has elapsed \\ 
  altart & The target is above the artificial horizon now \\ 
  altartmint & The target will still be above the artificial horizon after Min.Time has elapsed \\ 
  moonphase & The Moon is not too bright for the required observations \\ 
  moondist & The target is not too close to the Moon \\ 
  sunalt & The Sun is below Sun Alt now \\ 
  sunmintime & The Sun will still be below Sun Alt after Min.Time has elapsed \\ 
  flip & The telescope will not need to conduct a pier flip, or the job allows a pier flip to be conducted \\ 
 \hline  
\end{tabular}
\end{center}
\end{table}

Each time it is polled, the scheduler runs through every job in the queue and calculates a set of flags for each job, with a result of 0 indicating the flag is legal and 1 indicating it is illegal. An explanation of each of the flags is listed in Table \ref{tab:schedulerFlags}. The job is legal only if all the flags are 0. The scheduler then updates the priority of each job in the queue, as follows: 

\begin{itemize} 
\item If the job is defined as a Target of Opportunity (ToO=1 in Figure \ref{fig:jobfile}), its priority is given by the user-defined integer priority (Priority in Figure \ref{fig:jobfile}) plus airmass/100, where the airmass of the target is calculated at a point in time mid-way between the current time and the minimum end-time of the job (Min.Time in Figure \ref{fig:jobfile}). Hence a priority 2 job with a current airmass of 1.234 becomes 2.01234.
\item If the job is not a ToO (ToO=0 in Figure \ref{fig:jobfile}), its priority is given by the user-defined integer priority plus airmass/10. Hence a priority 2 job with a current airmass of 1.234 becomes priority 2.1234.
\item If any of the scheduler flags are non-zero, add 10 to the priority.  Hence an illegal, non ToO, priority 2 job with a current airmass of 1.234 becomes 12.1234.
\item If the priorities of two jobs are equal after the above updates, the job that entered the queue first has the higher priority.
\end{itemize}

The scheduler takes the job with the highest priority (equal to the lowest numerical value) in the queue and compares it with the updated priority of the job currently being observed. On the basis of this comparison, the scheduler decides what to do next, as follows:

\begin{enumerate}[\bf1.]
\item If the current observation is still the highest priority in the queue, and legal, continue observing.

\item If the current observation is no longer the highest priority in the queue, and the highest priority job in the queue is {\em not} a ToO, continue observing. This rule prevents jobs from constantly being overridden, and usually allows all of the requested observations to be obtained. Note that the exception to this rule is for priority 5 observations. These are ``queue fillers" and can be overridden immediately by any job in a higher-priority band, regardless of its ToO status. Queue filler jobs will also be resumed after interruptions, when conditions allow, which helps keep the telescope active even on quiet nights with few high priority jobs in the queue. 

\item If the current observation is no longer the highest priority in the queue, and the highest priority job {\em is} a ToO, interrupt the current observation by aborting the current exposure and immediately slewing to the new target. This rule guarantees that time-critical observations can be obtained. Note that if the current observation is also a ToO, it will not be interrupted unless the interrupting ToO lies in a higher priority band: for example, if the current observation is a ToO of priority 3.01234, it would not be overridden by a ToO job of priority 3.01000, but would be overridden by a ToO job of priority 2.01234.
\end{enumerate}

In this way the scheduler is simple and flexible. By setting appropriate values for the priority, ToO and start/stop times, it is possible to accommodate a wide variety of scientific projects, from high-priority time-critical jobs to low-priority long-term monitoring jobs. The drawback of the scheduler, as currently implemented, is that it does not look forward more than one job into the future. It therefore might observe a short high-priority job at the expense of a longer job with only a slightly lower priority. One solution would be to look forward to the end of the night and construct a schedule based on minimising some statistic, e.g. $\Sigma_{n}(t_{obs} \times p)$, where $t_{obs}$ is the length of time spent observing a job of priority $p$, summed over all of the jobs observed, $n$. The drawback of this approach is that we would then lose some control over which jobs are observed, and for this reason we have decided not to implement it.

\subsection{Conditions Monitor}\label{sec:conditionsmonitor}
The Conditions Monitor is a Python script that reads the weather information provided by up to three on-line sources from the ING, and is updated every 60 seconds. These sources record the meteorological conditions at three different locations at the Observatorio del Roque de los Muchachos: the weather masts at the JKT, Isaac Newton Telescope (INT) and WHT. The various parameters that are read from each mast are listed in Table \ref{tab:weather}, along with the limits defining when the dome should be closed. In order to prevent repeated opening and closing of the dome when conditions are close to any of the limits, we define different limits for when the dome should be re-opened following a closure due to bad weather. These re-opening limits are also listed in Table \ref{tab:weather}. 

\begin{table}
\begin{center}
\caption{Weather parameters considered by the Conditions Monitor, with different limits applicable depending on whether the dome is open or closed. The parameters are more lenient when the dome is already open compared to when the dome is closed, as the latter is likely to be because of recent bad weather, and some additional contingency is needed. This also helps prevent the dome from repeatedly opening and closing when conditions oscillate around the limits. The final two parameters (Max Sun and Max Moon) do not affect the opening and closing of the dome, but dictate when observing can commence.}
\label{tab:weather}
\begin{tabular}{p{5.78cm} p{0.85cm} p{1.0cm}}
  \hline
  \textbf{Parameter} & \textbf{Limit \newline(open)} & \textbf{Limit \newline(closed)}  \\ \hline
  Windspeed (m/s) & 10.0 & 9.0 \\ 
  Windspeed (km/h) & 36.0 & 32.4 \\ 
  Relative humidity (\%) & 70 & 63 \\ 
  Internal humidity (\%) & 70 & 63 \\ 
  Dry/Wet sensor & DRY & DRY \\
  Temperature (Celsius) & 1.0 & 2.0 \\ 
  Age of weather data (seconds) & 700 & 700 \\ 
  Max time for parameters to be unchanged (seconds) & 1200 & 1200 \\ 
  Max time without contact to Durham server (seconds) & 300 & 30 \\ 
  Max Sun altitude ($^{\circ}$) & -10  & -10 \\ 
  Max Moon altitude for dark time ($^{\circ}$) & -5 & -5 \\ 
 \hline  
\end{tabular}
\end{center}
\end{table}

The Conditions Monitor usually takes its input from the JKT mast. This is the most exposed location on the edge of the Caldera on La Palma, and hence usually reports the most extreme meteorological parameters. By using these parameters we are therefore taking the most conservative approach to opening and closing the dome, which is essential for a robotic telescope.

Occasionally there is a problem with the JKT mast, and one or more of the meteorological parameters reported by the system is either unavailable, corrupted or not updated. In this case, the Conditions Monitor automatically switches to taking its input from one of the other masts. If this mast is also in error, the Conditions Monitor will shut the dome. The Conditions Monitor also continuously checks that the internet link between \textit{pt5m} and the UK is working. If the link is broken, the Conditions Monitor will shut the dome. Once again, this is adopting a conservative approach: with no internet link, there is no possibility of us logging in remotely to shut the dome in case of a
problem. The system also notes the timestamps of all weather data packets, and treats the data as unreliable if it is more than 700 seconds old, or if the values don't change for more than 20 minutes.

The Conditions Monitor works by setting flags. All of the parameters listed in Table \ref{tab:weather} have an associated flag, which is set to 0 when the parameter is legal (i.e. it is safe to open the dome) and 1 when the parameter is illegal (i.e. it is not safe to open the dome). Only when all of the flags are 0 can the dome be opened, and if any of the flags are subsequently set to 1, the dome will close. When observing locally, and when sure that it is safe to do so, it is possible to override this system and allow the dome to remain open even if one or more of the flags are set to 1.

We have analysed the fraction of time that the dome remains open during night time hours, given the current weather parameters. Between 2013-01-20 and 2015-09-17 (971 nights), 21\% of night time experienced high humidity, 17\% experienced high windspeeds, and 10\% experienced low temperatures. These fractions will overlap with each other, as often high humidity is coupled with high winds and/or low temperatures. Overall, 32\% of night time was lost because of adverse weather, while 5\% was lost due to incomplete or corrupted meteorological information. Were we to be more liberal with our weather limits, e.g. setting the humidity limit to 80\%, the windspeed limit to 11 m/s, and the temperature limit to 0$^{\circ}$C for both open and closed dome scenarios, we could expect the fractions of lost night time to drop to 14\%, 11\% and 6\% for humidity, windspeed and temperature respectively. Despite this, we choose to maintain our current weather limits, as the risk of equipment damage outweighs the marginal gains made in observing time.

\subsection{Observation Database}
A Structured Query Language (SQL) database of all jobs submitted to \textit{pt5m} is hosted at the University of Sheffield, thus allowing observation requests to be submitted whilst the telescope is offline. Tables in the database include a list of {\em pointings}, the {\em exposures} associated with each pointing, and an {\em observing log}, which keeps track of the science images associated with each observation.

Submission of jobs to the SQL database is performed either via a web form\footnote{\url{http://slittlefair.staff.shef.ac.uk/pt5m/pt5mQueue.html}} which interfaces to the database through {\sc php} scripts, or by custom-written Python scripts which can be run anywhere, and interface to the same {\sc php} scripts. The {\sc php} scripts send the job details to the SQL database, which writes the ASCII files (Figure \ref{fig:jobfile}) to be picked up by the scheduler. Access to the database requires a username and password, which we administer directly to collaborators. 

A flag in the database keeps track of the status of the pointing: this flag can be set to indicate that observations are pending, completed, that the pointing has been deleted, the observations were interrupted, or the pointing's validity expired before it was observed. The pointing ID is used as a key to identify pointings in the {\em exposures} and {\em observing log} tables. The observing log has an entry for each science frame obtained with the telescope, including relevant information from the FITS headers.

\subsection{Alert Listener}\label{sec:followupSoftware}
In order to follow up the fastest-decaying transients, the human processing stage of an alert system needs to be bypassed, as this can lose valuable time. The Alert Listener was developed specifically to perform transient follow-up, and is composed of three elements: software to listen for alerts and schedule observations, software to automatically reduce the data, and software to publish the data. 

The listening software monitors transient alerts via \url{skyalert.org} and VOEvents \citep{seaman08}. It then analyses the alert information, and decides whether or not the event is worthy of follow-up observations with \textit{pt5m}. Currently, new cataclysmic variables, new AGN-variability, and new unknown transients are followed up, as well as any new Gamma Ray Bursts which have also triggered immediate radio follow-up with the AMI Large Array radio telescope \citep{staley13}. Transients classified as anything else are not followed up at this stage. If follow-up is requested, the software dictates what observations are required (e.g. a light curve for new cataclysmic variables, or BVRI colours for unclassified objects) and, using the magnitude given in the alert packet (if available), sets an appropriate exposure time. 

The data reduction component of the Alert Listener is a software daemon (Figure \ref{fig:software}) that automatically processes every science frame as soon as it has been read out. The following tasks are performed: 
\begin{itemize}
\item Debias and flat field using the master calibration frames (Section \ref{sec:pilot}). If no master frames are available from that night, the daemon searches backwards through previous nights until a master frame is found, up to a maximum look-back time of 30 nights. 
\item Using the {\sc SExtractor} package, extract a catalogue of all stellar sources in the image.
\item Match the extracted catalogue to published \textit{Vizier} catalogues and conduct an astrometric calibration, allowing for rotation, shift and stretch. 
\end{itemize}
The daemon also records this activity in a nightly log, including any failures to complete the tasks.

The third component of the Alert Listener automatically publishes the transient data. Each extracted catalogue associated with a transient alert is automatically uploaded to the Gaia Science Alerts Calibration Server\footnote{\url{http://gsaweb.ast.cam.ac.uk/followup}} which calculates and publishes a calibrated magnitude for the transient. We use this server because we have been providing photometric follow-up for the Gaia Science Alerts project (see e.g. \citealt{campbell15b}), and it is a useful repository for transient data.

\subsection{Autoguiding}\label{sec:autoguiding}
Since \textit{pt5m} does not have a separate autoguider, we have implemented autoguiding using the science frames. For each new image, the catalogue of sources extracted by the data reduction daemon (Section \ref{sec:followupSoftware}) is compared with that of the first science frame in the sequence. A minimum of 11 stars are needed for an accurate comparison; any fewer than this and the autoguiding process is skipped for that image. The median offset between the catalogues is then used to adjust the telescope position. The offset is applied during the readout of the following image. Guiding can be enabled or disabled at any time and is automatically re-established following a pier flip. The downside to our autoguiding implementation is that when particularly long exposures are needed, the guiding is unable to correct for tracking errors on shorter timescales than the exposure time. Equally, when very short cadence observations are required ($<10$ seconds including exposure time and read out), the data reduction process can sometimes cause minor delays. However, observations of such short cadence are very rare. Section \ref{sec:tracking} describes the performance of the telescope tracking with and without autoguiding enabled. 


\section{Performance}\label{sec:comms}
All of the performance data given in this section refer to the final configuration of the telescope described in Section \ref{sec:history}.

\subsection{Detector}
The CCD readout noise, gain and flat field noise were measured from bias frames and dome flats using the photon transfer curve technique \citep{janesick01} and are shown in Table \ref{tab:parameters}. The gain is unchanged when binning, but the readout noise depends on the binning factor. The flat field noise is the same in all filters, to within the measurement uncertainties, suggesting the dominant source of this noise is variations in the physical size of the pixels, rather than some wavelength dependent response. 

The dark current was measured at night with the dome, mirror and camera shutters closed. Exposure times of 20 minutes were used, and the dark current was measured as the median count level after the median bias level (measured only minutes earlier) had been subtracted. The detector shows negligible dark current ($8\pm4$ e$^{-}$/pixel/hour), but this tends to be concentrated in hot pixels and hence 
it can be important to use median-combined dark frames when reducing data.

By observing a star field with a range of exposure times in quick succession, we were able to analyse the linearity of the detector. We found that the detector performs linearly to within 0.5\% at light levels of up to 45000 counts/pixel, but departs from linearity by up to 2\% at higher light levels. 

\subsection{Optics}
To measure the performance of the telescope optics, we observed a star cluster filling the entire field of view in all five available filters ($H\alpha$, $B$, $V$, $R$, $I$). We measured the full width at half maximum (FWHM) of stars across the image. Our best measurements had a FWHM of 1.3". This is worse than expected given the median seeing at the site of 0.69" \citep{wilson99} and the intrinsic optical performance of the OTA (which is close to diffraction limited at optical wavelengths, according to the manufacturer's tests). The most likely contributions to this discrepancy are the local seeing due to the roof-top location of \textit{pt5m}, and the effects of wind shake.

\begin{figure}
\centering
 \includegraphics[width=0.49\textwidth]{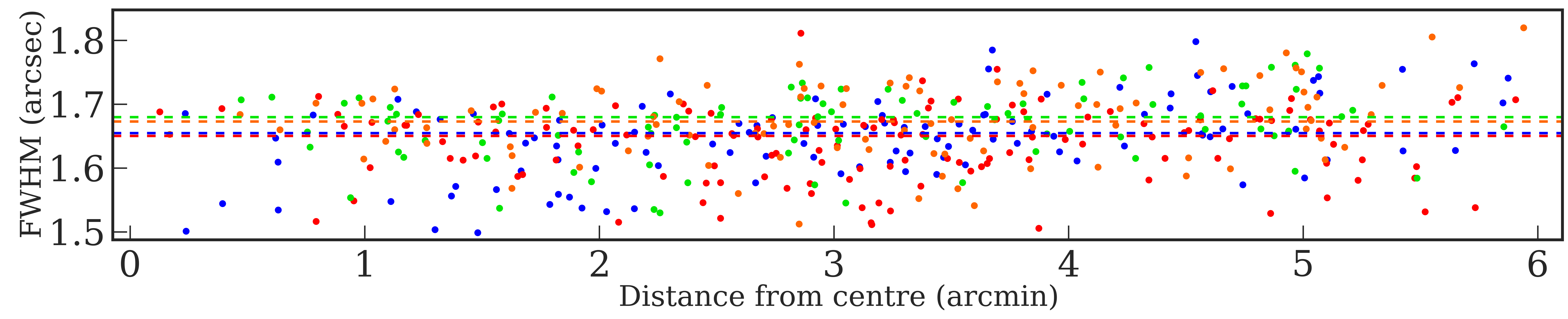}
 \caption{FWHM in the $R$-band as a function of radial distance from the centre of the CCD, measured from an image of the open cluster IC 4996. The standard deviation of the FWHM is 0.06" in this image. The four colours represent four different quadrants of the CCD; blue is top right, red is bottom right, green is top left and orange is bottom left. The horizontal lines represent the median FWHM in each quadrant.}
 \label{fig:imquality}
\end{figure}

The image quality appears to be largely independent of wavelength; the difference in FWHM from filter to filter is less than 10\% (approximately $2 \sigma$). The image quality degrades by approximately 0.05" from the centre to the corners of the CCD, as shown in Figure \ref{fig:imquality}, in accordance with the expectation from ray tracing of the telescope optics.

Flat-field images show signs of vignetting, with the drop in flux from the bright central region to the darker corners of a typical sky-flat being 10-15\%. The vignetting pattern is centred slightly above chip centre, suggesting that the science camera is mounted just off centre of the optical axis. 

\subsection{Zeropoints}
The photometric zeropoints in the four broad-band filters were measured using the standard stars SA101-315 and SA102-620. Each star was observed twice on two separate photometric nights, at different airmasses. We corrected for extinction using the coefficients given in the ING Observers Manual, 1995\footnote{\url{http://www.ing.iac.es/astronomy/observing/manuals/ps/general/obs_guide.pdf}}. The above-atmosphere zeropoints representing the magnitude of a star giving one photo-electron per second in each filter were then calculated. The final zeropoints, representing the mean of all four standard star measurements in each filter, are shown in Table \ref{tab:parameters}.

\begin{figure}
\centering
 \includegraphics[width=6.5cm,angle=270]{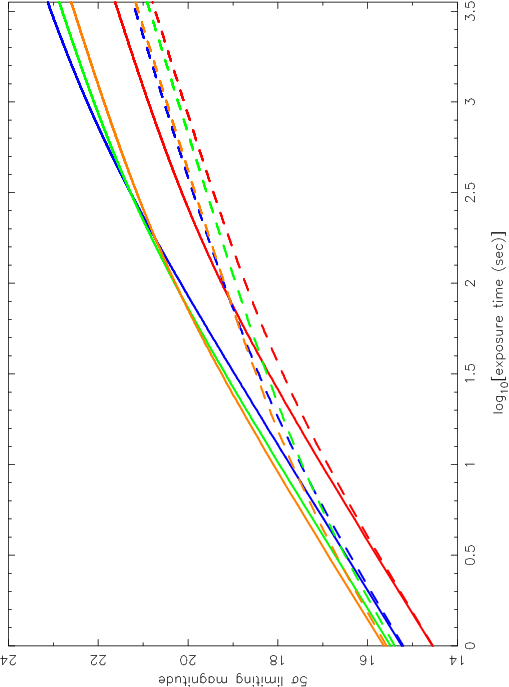}
 \caption{$5\sigma$ limiting magnitudes of \textit{pt5m} as a function of exposure time. The blue, green, orange and red curves show the results for the $B$, $V$, $R$ and $I$ filters respectively. Solid lines show the results for dark time and dashed lines for bright time. The calculations assume seeing of 1.5", airmass of 1.0 and 2x2 binning.}
 \label{fig:snr}
\end{figure}

Figure \ref{fig:snr} shows the $5\sigma$ limiting magnitudes that \textit{pt5m} can achieve as a function of exposure time and moon brightness, calculated using the measured zeropoints. 

\subsection{Pointing, Tracking and Guiding}\label{sec:tracking}
The pointing accuracy of \textit{pt5m} is 0.4' (RMS), with marginally worse pointing at low elevations and high declinations. The tracking accuracy of \textit{pt5m} was measured from observations following the same targets for 4 hours. The worst measured tracking drifts were up to 30"/hour for certain areas of the sky (at low elevations and at high declinations), but on average tracking drifts of 12"/hour were seen. This drift is larger than expected according to the mount specifications, most probably due to small errors in the mount's polar alignment, and flexure in the telescope and roof-top platform. The drift limits exposure times to a maximum of 5 minutes. With the introduction of autoguiding (Section \ref{sec:autoguiding}), positional fluctuations of only 1.3" (RMS) from the original position are seen.

\subsection{Response Times}
It takes less than two seconds from the announcement of a Gamma-Ray Burst as a VOEvent for the Alert Listener (Section \ref{sec:followupSoftware}) to submit a job for follow-up observation to the \textit{pt5m} queue. The pilot checks the scheduler for the highest priority job every 15 seconds. We estimate a maximum slew time of 30 seconds, assuming that the telescope has to conduct a pier flip and move to the opposite side of the sky. Adding these together we can expect to be tracking on target less than a minute after an alert is announced. Unfortunately we are yet to receive an observable night time trigger during safe weather conditions.

\subsection{Residual Images}
In October 2013, we serendipitously observed an apparent transient event whilst monitoring the active blazar 4C+38.41 \citep{gasparrini13}. The bright (\textit{B}$\sim$16) transient was observed in the same field, but was located several arcminutes away and was not thought to be associated with the blazar. The transient appeared to fade quickly (0.2 magnitudes/minute) over the course of around 15 minutes, and moved slowly (0.5"/min) with respect to background stars. We could not identify the transient as a known comet, asteroid or minor planet, and the coordinates did not coincide with the orbit of any known man-made satellite. The object appeared to have a stellar point spread function. Bizarrely, the transient reappeared two weeks later, 18" away from the original position, and this time appearing as a double source. 

After exhausting all possible astrophysical origins, we later discovered that this fast-fading transient signal was in fact the residual image of a bright star used for focusing just minutes earlier. Photo-electrons appear to have been trapped in impurity sites in the substrate of the chip. After the CCD was read out and cleared, these trapped electrons gradually leaked out during the next few exposures, creating a spurious signal in the resulting images. 

This residual image phenomenon is not uncommon in infrared detectors, but is rare in optical detectors \citep{rest02}. Our Kodak KAF-3200ME CCD is known to exhibit this behaviour\footnote{\url{http://canburytech.net/QSI532/RBI.html}}. Clearly, care must be taken when analysing images taken immediately after exposures containing bright stars.


\section{Example Science}\label{sec:results}
We used \textit{pt5m} to conduct an extensive study of the transiting exoplanet WASP-33b \citep{collier10}. This hot Jupiter is unusual in that it orbits an A-type star which itself exhibits $\delta$-Scuti pulsations. The planet is also observed to be hotter and more bloated than most other exoplanets. In addition, there have been tentative claims of a resonance between the period of orbit and period of pulsations, suggesting a tidal link \citep{herrero11}.

\begin{figure}
\centering
 \includegraphics[width=0.48\textwidth]{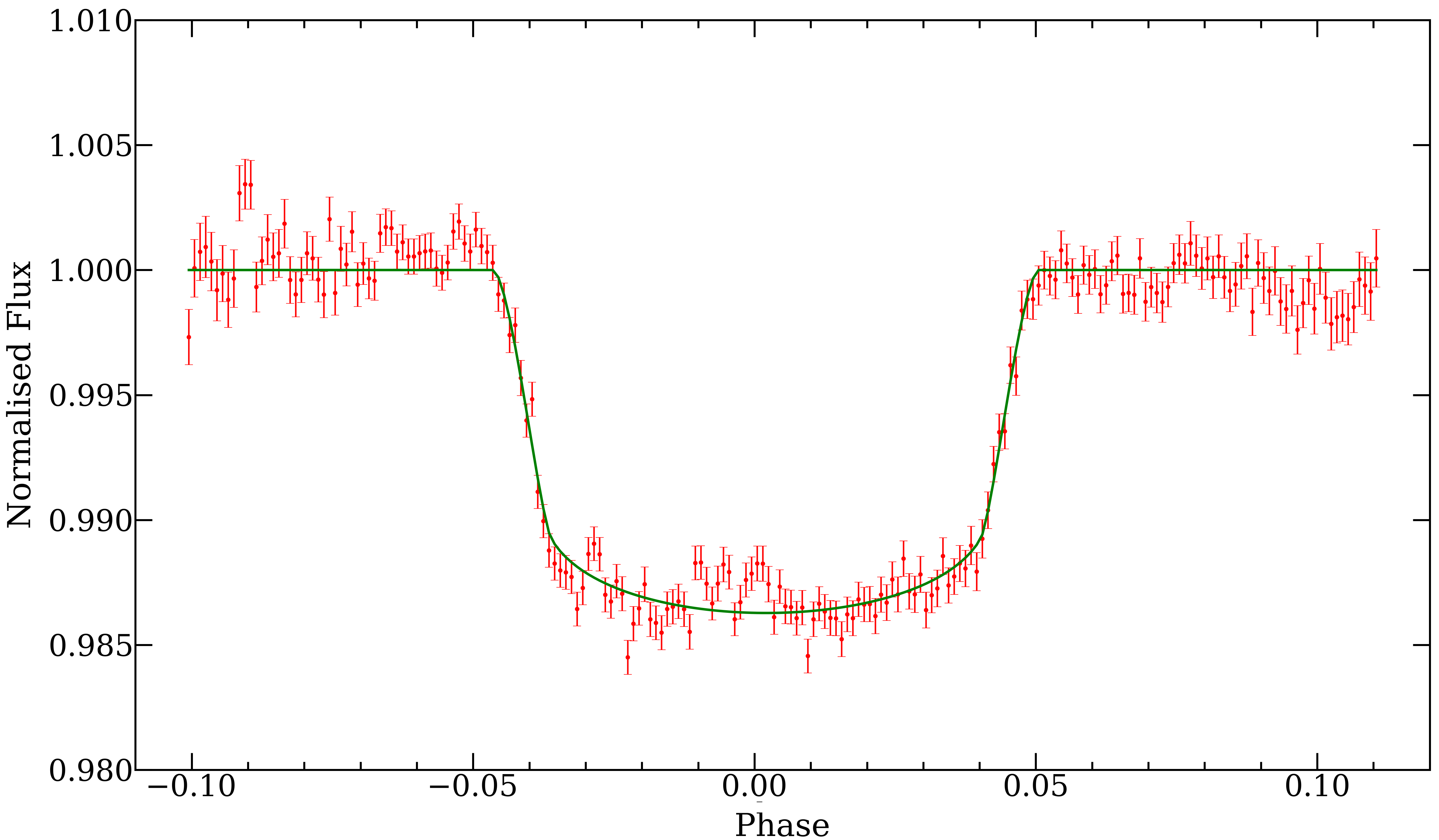}
 \caption{Transit light curve of WASP-33b, comprised of 15 individual transits. The observations have been phase-folded, binned and combined using inverse-variance weighting. The solid curve shows our best fit model. See text for details.}
 \label{fig:wasp33}
\end{figure}

We observed a total of 15 transits of WASP-33b in 2011 and 2012. We also collected several hours of out-of-transit data in order to study the stellar pulsations. The observations were phase folded using our own ephemeris, binned and combined using inverse-variance weighting. The combined transit light curve is shown in Figure \ref{fig:wasp33}. We model the planetary transit using the geometrical analysis of \citet{sackett99}, finding a planetary radius of $1.603\pm0.014$ $R_{J}$ and an orbital inclination of $87.04\pm0.26^{\circ}$. Uncertainties were estimated using the `prayer-bead' technique of residual permutation \citep{desert09}. 

\begin{figure}
\centering
 \includegraphics[width=0.48\textwidth]{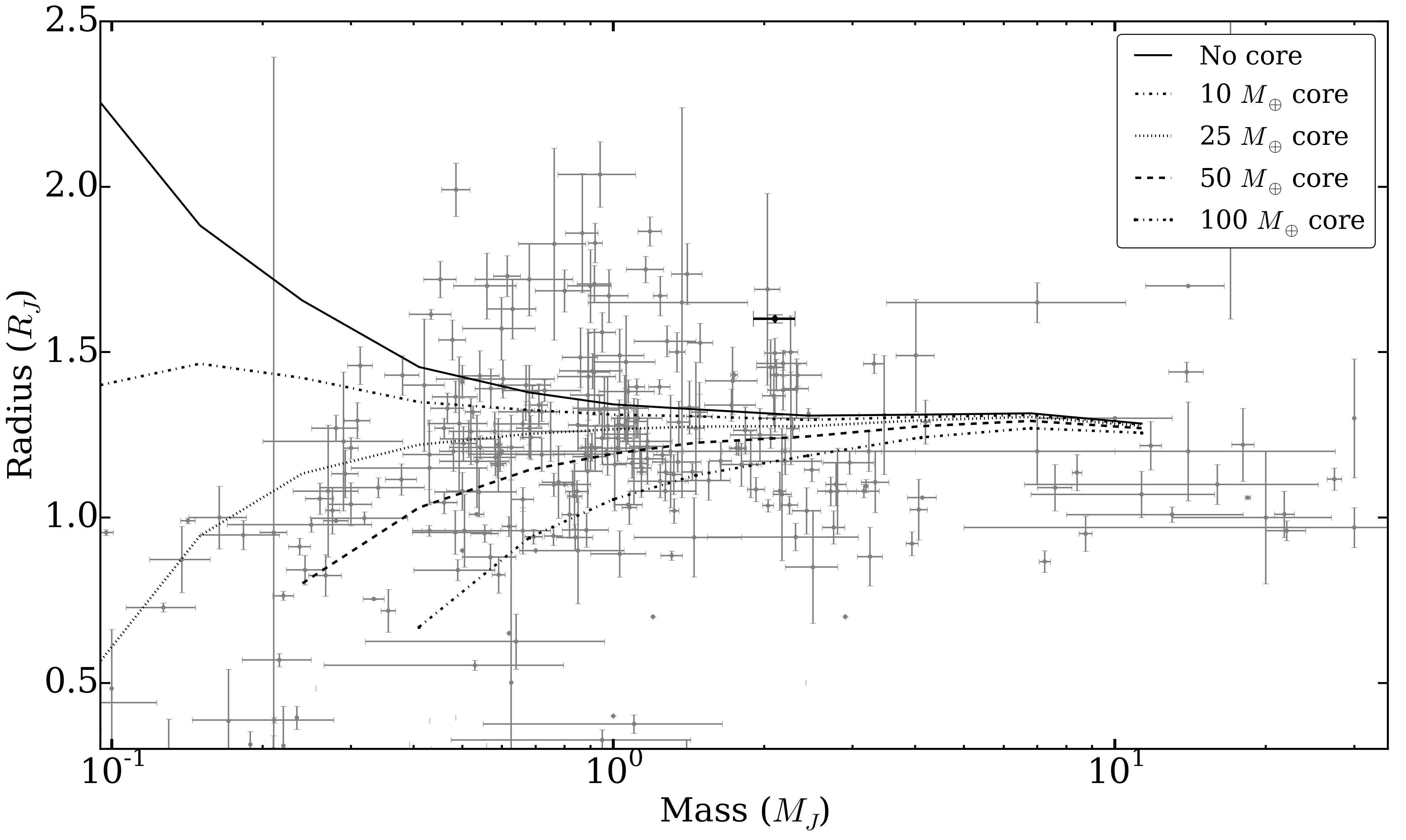}
 \caption{Mass-radius diagram showing the known exoplanets with measured or estimated masses and radii, as of July 2015 \citep{schneider11}, plotted as grey points. WASP-33b is shown in black, with the planetary mass measured by \citet{lehmann15}. Also shown are planetary models of a H/He planet of age 300 Myr and a scaled orbital separation of 0.02 au, with five different core masses \citep{fortney07}.}
 \label{fig:massradius}
\end{figure}

Our measured inclination matches well with previous studies \citep{collier10,herrero11,smith11,kovacs13,vonessen14}, but the planetary radius is found to be up to 3$\sigma$ larger. This highlights the fact that WASP-33b is substantially bloated in comparison to other exoplanets. Our measurement of the radius is plotted on a mass-radius diagram and compared to theoretical models of exoplanetary structure in Figure \ref{fig:massradius}. The measured radius is almost 20\% larger than expected from planetary models \citep{fortney07}, although it should be noted that these models are not tailored to such a close-in exoplanet as WASP-33b. 

The origin of an enlarged planetary radius could be attributed to a number of factors, including tidal heating \citep{bodenheimer01}, Ohmic heating \citep{batygin10} and kinetic heating \citep{guillot02}. The circular orbit of WASP-33b \citep{smith11,deming12} rules out tidal heating, but Ohmic and kinetic heating may play a role. However, we propose an additional and simpler reason for its enlarged radius; the young age of the system. The age of WASP-33 could be as young as 10 Myr \citep{moya11}, making it one of the youngest known exoplanet hosts. The size of an exoplanet is expected to decrease with time after formation, as the planet cools and contracts \citep{fortney07}. In addition, standard planetary models usually use arbitrary starting conditions, with their main scientific interest being the state of the planet some Gyrs after formation, by which time their starting conditions have little effect \citep{fortney10}. In this sense, it is not surprising that WASP-33b is found to be bloated with respect to these models, as it is much younger than the planets the models are designed to simulate. 

Our phase-folded light curve (Figure \ref{fig:wasp33}) shows a mid-transit bump, which is present in several of the individual transit light curves. The star spot explanation for this phenomenon proposed by \citet{kovacs13} is unlikely, as the host is an A-type star. Instead, we propose that the bump may be due to a pulsation in the host star resonating with the orbital period, and therefore is still visible after phase folding. However, \citet{vonessen14} find no conclusive evidence for such star-planet interactions, and we found no pulsation periods in resonance with the orbital period in power spectra of the out-of-transit light curves. The nature of the bump remains unclear.

\section{Conclusions}
We have described the design and performance of \textit{pt5m} -- a 0.5m robotic telescope on the roof of the WHT on La Palma. The facility is primarily used for research, and we 
present an example study of the transiting exoplanet WASP-33b performed with \textit{pt5m}. The facility also functions as a valuable teaching resource, and is used regularly for undergraduate projects at both the University of Sheffield and Durham University. Interested readers are encouraged to visit the \textit{pt5m} website\footnote{\url{https://sites.google.com/site/point5metre/}} for further information. 


\section*{Acknowledgments}
\textit{pt5m} is a collaborative effort between the Universities of Durham and Sheffield. The telescope is kindly hosted by the Isaac Newton Group of Telescopes, La Palma. The Roque de los Muchachos Observatory is operated by the Instituto de Astrof\`{i}sica de Canarias. Financial contributions from the University of Sheffield Alumni Foundation are gratefully acknowledged. We are also grateful to the Science and Technology Facilities Council for financial support in the form of grant ST/L00075X/1. LKH acknowledges support via the Harry Worthington Scholarship at the University of Sheffield, and the ING studentship programme. Raw data used in the analysis presented here are available from the lead author.

We thank the following staff at ING, Durham and Sheffield for providing technical support:  Neil O'Mahony, Juerg Rey, Alan Chopping, Kevin Dee, Marc Dubbeldam, Dora Fohring, Simon Blake, Steve Rolt, Trevor Gamble and Simon Dixon. We also thank Tim Staley for his {\sc voevent-parse} Python package.

\bibliographystyle{mn2e}
\bibliography{abbrev,refs}

\label{lastpage}

\end{document}